\documentclass[preprint, 1p]{elsarticle}

\usepackage{amssymb}
\usepackage{amsmath}
\usepackage{graphicx}

\usepackage[utf8]{inputenc}
\usepackage[T1]{fontenc}
\usepackage{etoolbox}
\usepackage[dvipsnames]{xcolor}
\usepackage{hyperref}
\usepackage{cleveref}
\usepackage{comment}
\usepackage{longtable}

\newcommand{\phavg}[1]{\langle#1 \rangle}
\newcommand{\favg}[1]{\tilde{#1}}
\newcommand{\wfavg}[1]{\widetilde{#1}}
\newcommand{\pavg}[1]{\langle #1 \rangle_\mathrm{p}}

\newcommand{\fu}{\favg{u}}
\newcommand{\Rij}{\wfavg{R}_{ij}}

\newcommand{\subpar}[1]{{#1}_{\parallel,\parallel}}
\newcommand{\subperp}[1]{{#1}_{\perp,\perp}}
\newcommand{\Rpar}{\subpar{\wfavg{R}}}
\newcommand{\Rperp}{\subperp{\wfavg{R}}}
\newcommand{\Rparp}{\Rpar^{\mathrm{p}}}
\newcommand{\Rperpp}{\Rperp^{\mathrm{p}}}
\newcommand{\bpar}{\subpar{b}}
\newcommand{\bperp}{\subperp{b}}
\newcommand{\kdivK}{\hat{k}}
\newcommand{\phrho}{\phavg{\rho}}

\newcommand{\Tu}{\wfavg{T}_{\parallel}}

\newcommand{\Tup}{\Tu^\mathrm{p}}

\newcommand{\fulldep}{\left(\alp,\Rep,\Map\right)}

\newcommand{\Rep}{\mathrm{Re}_\mathrm{p}}
\newcommand{\Rem}{\mathrm{Re}_\mathrm{m}}

\newcommand{\Map}{\mathrm{Ma}_\mathrm{p}}
\newcommand{\Cd}{C_\mathrm{D}}
\newcommand{\Dp}{D_\mathrm{p}}
\newcommand{\alp}{\alpha_\mathrm{p}}

\journal{International Journal of Multiphase Flow}

\begin{document} 

\begin{frontmatter}



\title{Pseudo-turbulence models for compressible flow through random particle suspensions}

\author[inst1]{Andreas Nygård Osnes}
\author[inst1]{Magnus Vartdal}

\affiliation[inst1]{organization={Norwegian Defence Research Establishment},
            addressline={PO Box 25}, 
            city={Kjeller},
            postcode={2007}, 
            country={Norway}}

\begin{abstract}
A model for the pseudo-turbulent Reynolds stress tensor in compressible flows through monodisperse particle clouds is developed based on data from particle resolved numerical simulations. This model extends previous models for the incompressible regime to subsonic bulk Mach numbers. A Lagrangian model for the local pseudo-turbulent Reynolds stress around a particle, based on the correlation of a local estimate of the Reynolds stress and the drag force on the particle, is also presented. This model can be employed in Euler-Lagrange type simulations. Additionally, corresponding models for the pseudo-turbulent transport of turbulence kinetic energy, the velocity triple-correlation in the volume averaged total energy equation, is also introduced as a first effort aimed at closing the energy equation.
\end{abstract}


\begin{highlights}
\item Models for the pseudo-turbulent velocity moments in the volume-averaged momentum and energy equations are developed for finite particle volume fraction, Reynolds number and Mach number.
\item Lagrangian forms of the pseudo-turbulent closure models are developed, which incorporate the correlation between drag and pseudo-turbulence. 

\end{highlights}


\end{frontmatter}


\section{Introduction}

Flows through particle suspensions are strongly inhomogeneous at the particle scale, and a bulk description of such systems is incomplete without considering the statistics of the particle scale fluctuations. The particle-scale fluctuations are referred to as pseudo-turbulent fluctuations \cite{mehrabadi2015}. Meso- and macroscale simulations of such flows must, in general, account for pseudo-turbulence, as it may be dynamically important. In this work, we present correlations for pseudo-turbulent Reynolds stresses and pseudo-turbulent transport of pseudo-turbulent kinetic energy, applicable for compressible flows through particle suspensions. We provide two forms of the models - a conventional Eulerian algebraic form, and a Lagrangian form where we associate the particle-induced fluid phase fluctuations with the particles.

At finite Mach number, local flow contraction/expansion can cause rapid changes in the flow state, which can have a substantial effect on the interphase momentum and energy transfer, mixing characteristics, chemical reaction rates, etc. Models that capture these compressibility effects, and can be applied in meso/macroscale simulations, are crucial for development of technological applications of multiphase flows, as well as for understanding the vast number of natural phenomena in which multiphase flows play an important role. Examples include spacecraft landing \citep{capecelatro2022}, jet noise suppression \citep{krothapalli2003}, fuel injection systems \citep{samareh2008} and pyroclastic currents \citep{valentine2018}.

Across engineering applications and scientific studies, simulations are performed at various scales, which necessitates the development of appropriate closure models for each scale. When the information on the exact distribution of particles is available, techniques for reconstructing the forces on individual particles \citep{akiki2016} as well as particle scale fluid field fluctuations \citep{moore2019,siddani2021a} have been developed for the incompressible regime, and similar methods may potentially be used for compressible flows. For macroscale simulations, where the exact distribution of particles is unknown or the number of particles is too large, the abovementioned techniques are inapplicable and alternative modeling approaches must be used. These latter simulations require computationally inexpensive closures, which motivates the development of algebraic or stochastic models.

The most dynamically important of the unclosed terms in the macroscale fluid phase equations is the pseudo-turbulent Reynolds stress tensor, $\Rij$.  \citet{mehrabadi2015} presented an algebraic model for $\Rij$ in the incompressible regime, but for finite Mach numbers, the only algebraic model available is the relatively crude approximation of \citet{osnes2019}. An alternative approach is to introduce additional transport equations for the pseudo-turbulence, as presented for compressible flows in \cite{vartdal2018using, shallcross2020}. However, these equations introduce their own closure problems. The importance of including models for $\Rij$ has been demonstrated for simulations of shocked particle curtains \cite{shallcross2020, osnes2021}, and we expect the same to hold for other compressible particle-laden flows.

The goal of the present paper is to develop algebraic models for the pseudo-turbulent Reynolds stresses that capture the effects of volume fraction, Reynolds number, and Mach number. We construct models for the mean value as well as a Lagrangian model for the variation in pseudo-turbulent Reynolds stresses that correlates with drag on individual particles. The model for the mean pseudo-turbulent Reynolds stresses reduces to the model of \citet{mehrabadi2015} in the incompressible limit. Furthermore, we present corresponding models for the pseudo-turbulent transport of pseudo-turbulent kinetic energy, i.e. a triple-velocity correlation, which appears in the total energy equation.

The present model development is based on flow data from particle-resolved simulations of statistically steady flows through monodisperse, random, fixed particle beds.  Particle-resolved simulation is a well-established tool for study of gas-solid flows, which has been widely used in both the compressible and incompressible regimes; see for instance \citep{tenneti2011,akiki2016,sen2018,osnes2019}.

\section{Problem set-up}
The datasets used for model development in this work were produced with particle-resolved simulations, and have previously been used to develop models for particle forces \cite{osnes2023}. The computational domains are triply periodic boxes filled with random distributions of fixed rigid particles, occupying a mean volume fraction, $\alp$. The dimensions of the boxes are $30\Dp\times10\Dp\times10\Dp$ for $\alp\leq 0.2$, and $15\Dp\times10\Dp\times10\Dp$ for $\alp=0.3$, where $\Dp$ is the diameter of a particle. These domain sizes are larger than those suggested by \citet{elmestikawy2025influence} to avoid excessive correlations of the forces on the particles. 

The governing equations of the fluid flow between the particles are the compressible Navier--Stokes equations. The equations are closed using a perfect gas equation of state, with an adiabatic index of 1.4, along with a power-law dependence of the viscosity, with an exponent of 0.75, and a constant Prandtl number of 0.7. The boundary conditions on the particle surfaces are isothermal, no-slip, and no-penetration conditions. The fluid flow is driven by imposing a constant acceleration in the $x$-direction. The resulting equations were solved using the compressible flow solver CharLES~\citep{bres2017}.

The simulations were run until a steady-state flow was achieved, at which point the particle forces and heat flux across the particle surfaces balance the imposed force and the work done by it. The mean flow through the system can be described in terms of the particle volume fraction, Reynolds number, and Mach number
\begin{equation}
    \alp=\frac{N_p\pi D_p^3}{6 V} \, , \quad \Rep=\frac{\phrho\fu\Dp}{\phavg{\mu}}, \quad \Map=\frac{\favg{u}}{\phavg{c}},
    \label{eq:RepMap}
\end{equation}
where $N_p$ is the number of particles, $V$ is the domain volume, $c$ is the speed of sound, $u$ the mean streamwise velocity, and $\rho$ and $\mu$ are the fluid density and dynamic viscosity, respectively. Here, $\phavg{\cdot}$ denotes the average of a quantity over the volume occupied by the fluid phase, and $\favg{\cdot}=\phavg{\rho \cdot}/\phavg{\rho}$ denotes the corresponding Favre-average. 

The dataset consists of 42 simulations from the range $\alp \in [0.05, 0.3]$, $\Map \in [0.1, 0.87]$ and $\Rep \in [30,266]$. However, since the relation between the forcing function and the resulting Reynolds and Mach numbers is unknown, the resulting sampling of the parameter space is irregular. A table containing all the simulated configurations is found in \cite{osnes2023}.  

\begin{figure*}
    \centerline{
    \includegraphics{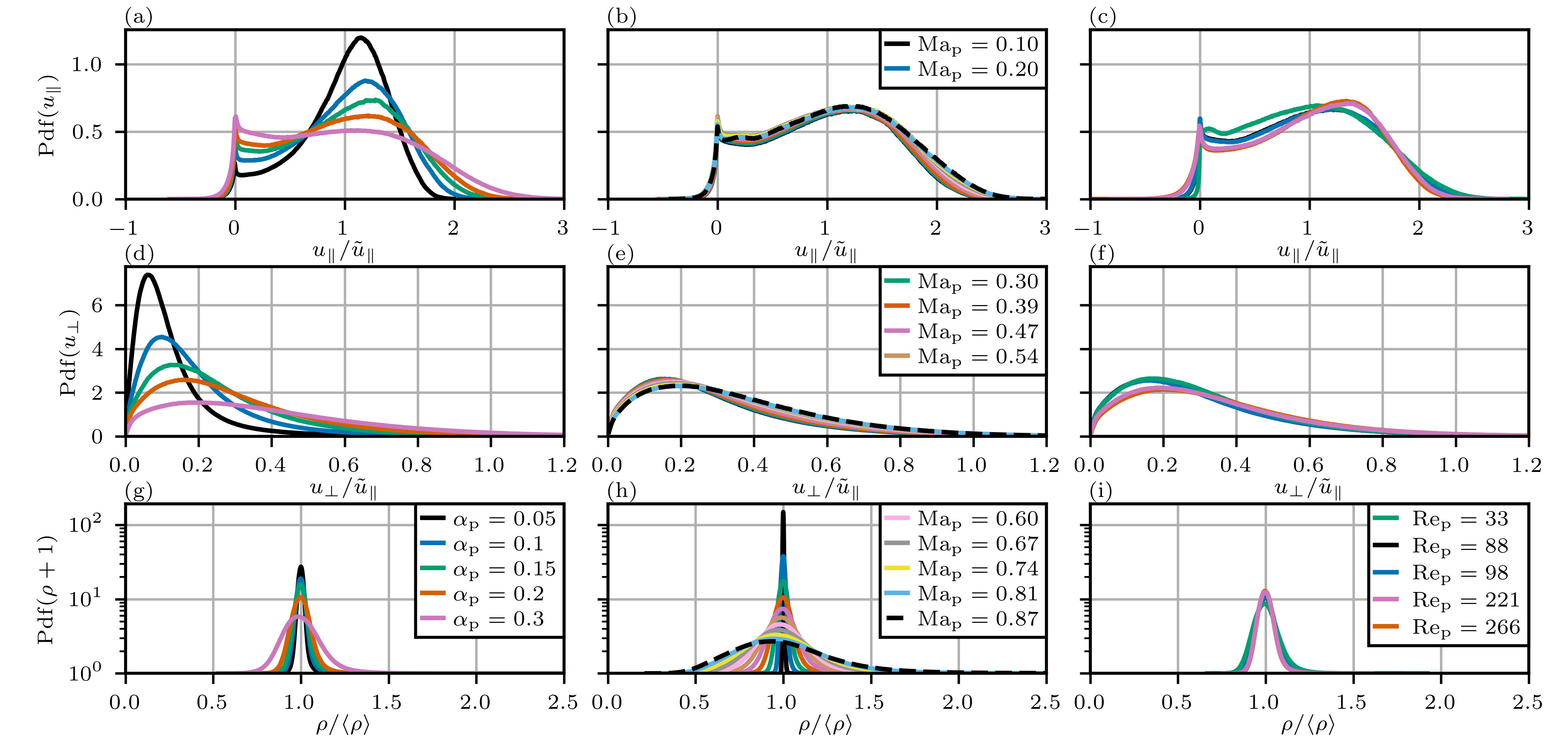}}
    \caption{Distribution of streamwise velocity, (a, b, c), spanwise velocity, (d, e, f), and density, (g, h, i) at $\Rep\approx100$, $\Map\approx0.4$ for varying $\alp$ (a, d, e), at $\alp=0.2$, $\Rep\approx100$ for different $\Map$ (b, e, h), and at $\alp=0.2$, $\Map\approx0.4$, for different $\Rep$ (c, f, i). Note that the $\Map$ legend has been split over (b, e, h), but is common for all three subfigures.}
    \label{fig:pdf_alphas}
\end{figure*}

In Euler-Euler and Euler-Lagrange simulations, the fluid flow is typically solved in a volume- or ensemble-averaged sense. These averaging operators introduce unclosed terms in the governing equations. In compressible particle-laden flows significant contributions to the momentum and energy equation from the pseudo-turbulent Reynolds stresses have been reported~\cite{sen2018,osnes2020}. Here, we seek to model this tensor, and to that end we compute the volume-averaged pseudo-turbulent Reynolds stress tensor 
\begin{equation}
\Rij=\widetilde{u_iu_j}-\favg{u_i}\favg{u_j}=\wfavg{u_i''u_j''},
\label{eq:RST}
\end{equation}
where $(\cdot)''$ denotes the fluctuation from the Favre-average and the Einstein summation notation is used, 
by integrating the steady-state simulation results over the entire computational domain. Due to the periodic nature of the domain, the tensor only has two independent components, i.e, the flow parallel ($\subpar{\cdot}$)  and orthogonal ($\subperp{\cdot}$) components. 
Furthermore, we also model the triple velocity correlation  
\begin{equation}
\wfavg{T}_i=\frac{1}{2}\widetilde{u_i''u_j''u_j''},
\label{eq:Tudef}
\end{equation}
which appears in the total energy equation. It represents the pseudo-turbulent transport of pseudo-turbulent kinetic energy. Due to the symmetries of the present configuration, $\wfavg{T}_i$ only has a non-zero component in the flow parallel direction, $\Tu$.

$\Rij$ and $\Tu$ are functions of the velocity and density distributions in the computational domain. To give an idea of the functional dependence of these distributions on the bulk flow parameters, a subset of the simulation results are shown in \cref{fig:pdf_alphas}. Among the three bulk flow parameters, the particle volume fraction has the largest impact on the distributions. The distributions are quite narrow at low $\alp$ but widen significantly as $\alp$ increases. In comparison, $\Rep$ has a moderate effect on the velocity distributions and an almost negligible effect on the density. Finally, for constant $\alp$ and $\Rep$, $\Map$ has a modest effect on the velocity distributions, and a major influence on the distribution of $\rho$.

\section{Eulerian models for \texorpdfstring{$\Rpar$}{Rpar}, \texorpdfstring{$\Rperp$}{Rperp} and \texorpdfstring{$\Tu$}{Tu}}
A model for the pseudo-turbulent Reynolds stress tensor at finite $\alp$ and $\Rep$ in the incompressible regime was proposed by \citet{mehrabadi2015}, and has been used by several authors also for compressible and non-isothermal flows \citep{peng2019implementation,balakrishnan2024}. The model comprises two algebraic equations; one for the ratio of pseudo-turbulent kinetic energy to mean flow kinetic energy, $\kdivK=k/K$, and one for the flow-parallel component of the pseudo-turbulent anisotropy tensor, $\bpar$. The flow-normal anisotropy component, $\bperp$, is related to $\bpar$ by $\bperp=-\bpar/2$. For homogeneous flows, this is a complete model for the pseudo-turbulent Reynolds stress tensor, $\Rij=2kb_{ij}+2k/3\delta_{ij}$. The two equations are

\begin{equation}
    \begin{split}
        \kdivK\left(\alp, \Rem\right) = 2\alp + 2.5\alp\left(1-\alp\right)^3\exp\left(-\alp\Rem^{1/2}\right),
    \end{split}
    \label{eq:kdivK_inc}
\end{equation}
\begin{equation}
    \bpar(\alp, \Rem) = \frac{0.523}{1+0.305\exp\left(-0.114\Rem\right)}\exp\left(\frac{-3.511\alp}{1+1.801\exp\left(-0.005\Rem\right)}\right),
    \label{eq:binc}
\end{equation}
where $\Rem = \left(1-\alp\right)\Rep$.  As shown in the previous section, the distributions of the velocity and density fields depend on $\Map$, and this is therefore also the case for the pseudo-turbulent Reynolds stress tensor. To account for the $\Map$-dependence, we propose extensions for \cref{eq:kdivK_inc,eq:binc} that incorporate the effect of finite $\Map$. The equations are multiplied by factors that are unity in the limit $\Map\rightarrow0$, and for higher $\Map$ captures the relative changes with increasing compressibility effects. The extended models can be written:

\begin{equation}
    \begin{split}
        \kdivK\left(\alp, \Rep, \Map\right) = \kdivK\left(\alp, \Rem\right)\left[1+\kdivK^{\Map}\fulldep\right],
    \end{split}
    \label{eq:kdivK}
\end{equation}
\begin{equation}
    \begin{split}
        \bpar\left(\alp, \Rep, \Map\right) = \bpar\left(\alp,\Rem\right)\left[1+\bpar^{\Map}\left(\alp,\Rep,\Map\right)\right],
    \end{split}
    \label{eq:b}
\end{equation}
where $\kdivK^{\Map}$ and $\bpar^{\Map}$ are functions that capture the effects of finite $\Map$. They are simple polynomials in terms of $\alp$ and $\Rep$, and includes a hyperbolic tangent term that captures the smooth increase of $\hat{k}$ and decrease of $\bpar$ with increasing $\Map$. Their full expressions are

\begin{equation}
    \begin{split}
        \kdivK^{\Map}\fulldep = \alp\left(C_1 + C_2\alp + \Rep^{C_3}\right)\left[\tanh\left(\frac{C_4}{C_5}\right) + \tanh\left(\frac{\Map-C_4}{C_5}\right)\right],
    \end{split}
    \label{eq:kdivK_map}
\end{equation}
\begin{equation}
    \begin{split}
        \bpar^{\Map}\fulldep = &\left[D_1 + \frac{\Rep}{300}\left(D_2 + D_3\frac{\Rep}{300}\right) + \alp\left(D_4 + D_5\frac{\Rep^2}{300^2} + D_6\alp\right)\right]\\
        &\times\left[\tanh\left(-\frac{D_7}{D_8}\right)-\tanh{\left(\frac{\Map-D_7}{D_8}\right)}\right].
    \end{split}
    \label{eq:bparmap}
\end{equation}
The coefficients $C_1, C2, \dots, C_5, D_1, D_2\dots D_8$ are given in \cref{tab:kdivk_coefs}. 

\begin{figure*}
    \centerline{
    \includegraphics{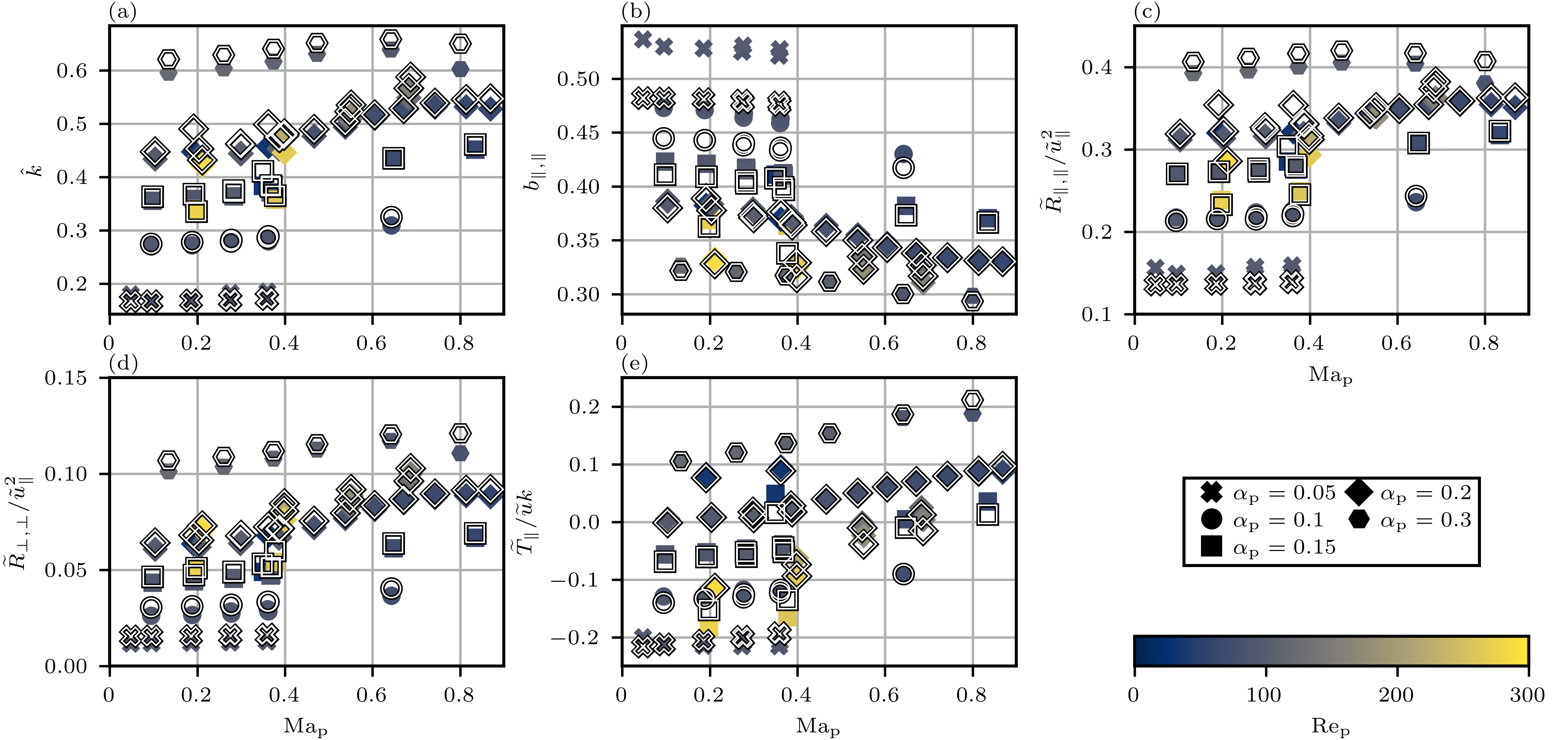}}
    \caption{Pseudo-turbulence fields as functions of $\Map$. (a): Ratio of pseudo-turbulent kinetic energy to mean flow kinetic energy. (b): Pseudo-turbulent anisotropy tensor. (c): Flow-parallel Reynolds stress. (d): Flow-normal Reynolds stress. (e): Velocity triple correlation. The particle-resolved simulation data are shown as filled symbols, and \cref{eq:kdivK,eq:kdivK_inc,eq:kdivK_map,eq:kdivK_inc,eq:kdivK_map,eq:b,eq:binc,eq:bparmap,eq:Tu} are shown with empty symbols.}
    \label{fig:kdivK_and_b}
\end{figure*}
\begin{table}
    \caption{Coefficients for the Mach-number extended pseudo-turbulent kinetic energy model, anisotropy model, and pseudo-turbulent transport of pseudo-turbulent kinetic energy model, \cref{eq:kdivK,eq:kdivK_inc,eq:kdivK_map,eq:b,eq:binc,eq:bparmap,eq:Tu}.}
    \label{tab:kdivk_coefs}
\begin{tabular}{c|c|c|c|c|c}
    Parameter &  Value & Parameter & Value & Parameter & Value\\
    \hline
    $C_1$ & -1.2152 & $D_1$ & -0.0462 & $E_1$ & -0.2906\\
    $C_2$ & -7.6314 & $D_2$ & -0.1068 & $E_2$ & 1.1899\\
    $C_3$ & 0.2889 &  $D_3$ & 0.6793 & $E_3$ &  0.5218\\
    $C_4$ & 0.6143 & $D_4$ & 1.1461 & $E_4$ &  0.0699\\
    $C_5$ & 0.3082 &$D_5$ & -2.6886 & & \\
     & &           $D_6$ & -2.1376 & & \\
     & &           $D_7$ & 0.4873 & & \\
     & &           $D_8$ & 0.2395 & & 
\end{tabular}
\end{table}
The simulation data and \cref{eq:kdivK,eq:b} are shown in \cref{fig:kdivK_and_b}. The ratio of pseudo-turbulent to mean kinetic energy increases with volume fraction, and also with $\Map$. The trends with $\Map$ are slightly different depending on $\alp$ and $\Rep$, but as the agreement between the empty and solid symbols in the figure demonstrates, the model can capture the varying trends. We observe that anisotropy decreases as $\Map$ increases, and note the anisotropy is underpredicted by the model at low volume fractions due to a disagreement between \cref{eq:binc} and the simulation data. This may be due to the relatively small domain size used in the computations on which \cref{eq:binc} is based.

We find that an appropriate normalization for $\Tu$ is to divide it by the product of the mean velocity and the pseudo-turbulent kinetic energy. We propose the following model

\begin{equation}
\frac{\Tu}{\fu k}\fulldep = E_1 + \frac{E_2\alp}{E_3 + \Rep/300} + E_4\Map.
\label{eq:Tu}
\end{equation}
The coefficients $E_1, E_2, E_3, E_4$ are given in \cref{tab:kdivk_coefs}. \Cref{eq:Tu} is compared against the simulation data in \cref{fig:kdivK_and_b} (e). $\Tu$ is overall an increasing function of $\Map$ and $\alp$. 

\section{Lagrangian models for  \texorpdfstring{$\Rpar$}{Rpar}, \texorpdfstring{$\Rperp$}{Rperp}, and \texorpdfstring{$\Tu$}{Tu}}

By definition, pseudo-turbulence is determined by the particle distribution and the particle motion relative to the fluid \cite{mehrabadi2015}. Therefore, a natural approach to modeling it is to assign pseudo-turbulence values to each particle. In an Eulerian-Lagrangian modeling setting one can then compute the Eulerian pseudo-turbulence fields by combining the pseudo-turbulence associated with all particles with some aggregation scheme, similar to how particle forces are transferred from the particle phase to the fluid phase, see e.g. \cite{capecelatro2013euler}.

The simplest way to assign pseudo-turbulence to each particle is to assume that there is no particle to particle variation at constant bulk flow properties, and that the entire Eulerian field can be associated to the particles. Then, the particle-associated pseudo-turbulence values are simply the Eulerian values divided by the particle number density $n=N_p/V$. This simple approach still presents a more sophisticated model for pseudo-turbulence than the Eulerian models in \cref{eq:kdivK,eq:b} because upon aggregation to the Eulerian grid, the local particle configuration affects the resulting Eulerian pseudo-turbulence field. 

The fluid forces acting on a particle depend on the flow in its immediate vicinity, and are therefore inherently coupled to the pseudo-turbulence. Furthermore, one of the primary production terms of pseudo-turbulence is proportional to the force on the particles and the relative velocity \cite{vartdal2018using}, thus we can expect some level of correlation between the pseudo-turbulent fields in limited volumes around each particle and the particle forces. Here, we propose a model that takes advantage of this correlation to model the particle-to-particle variation of the pseudo-turbulence. This approach relies on a model for $\Cd'$, the difference between the drag coefficient of an individual particle and the mean $\Cd$ at given bulk flow conditions. Such models are, for instance, available in \citet{lattanzi2020stochastic} and \citet{osnes2023}.

We define a spherical volume around each particle which we assume to be associated with the particle. For each particle we take the Favre-average of $u_\parallel''u_\parallel''$ and $u_\perp''u_\perp''$ and $u_\parallel''u_i''u_i''$ over that volume (note that fluctuations are still computed relative to the average over the full computational domain) to form particle-associated values of $\Rpar$, $\Rperp$, and $\Tu$, which we will denote $\Rparp$, $\Rperpp$, and $\Tup$, respectively. We choose spherical volumes with diameters $3\Dp$. This corresponds to the width of the volume filtering kernel recommended by \citet{capecelatro2013euler} for transferring particle-data to the Eulerian mesh. The averages of $\Rparp$, $\Rperpp$ and $\Tup$ are close to $\Rpar$, $\Rperp$ and $\Tu$ with this choice, except for $\Tup$ at $\alp=0.05$. Note that, depending on the particle configuration, this approach leads to overlapping particle-associated volumes, and additionally, parts of the fluid volume can be unaccounted for if the local particle volume fraction is low. We seek to model the distributions of $\Rparp$, $\Rperpp$ and $\Tup$. These distribution models can be used in conjunction with \cref{eq:kdivK,eq:b,eq:Tu} to compute the particle-associated pseudo-turbulence fields. 

\begin{figure*}
    \centerline{
    \includegraphics{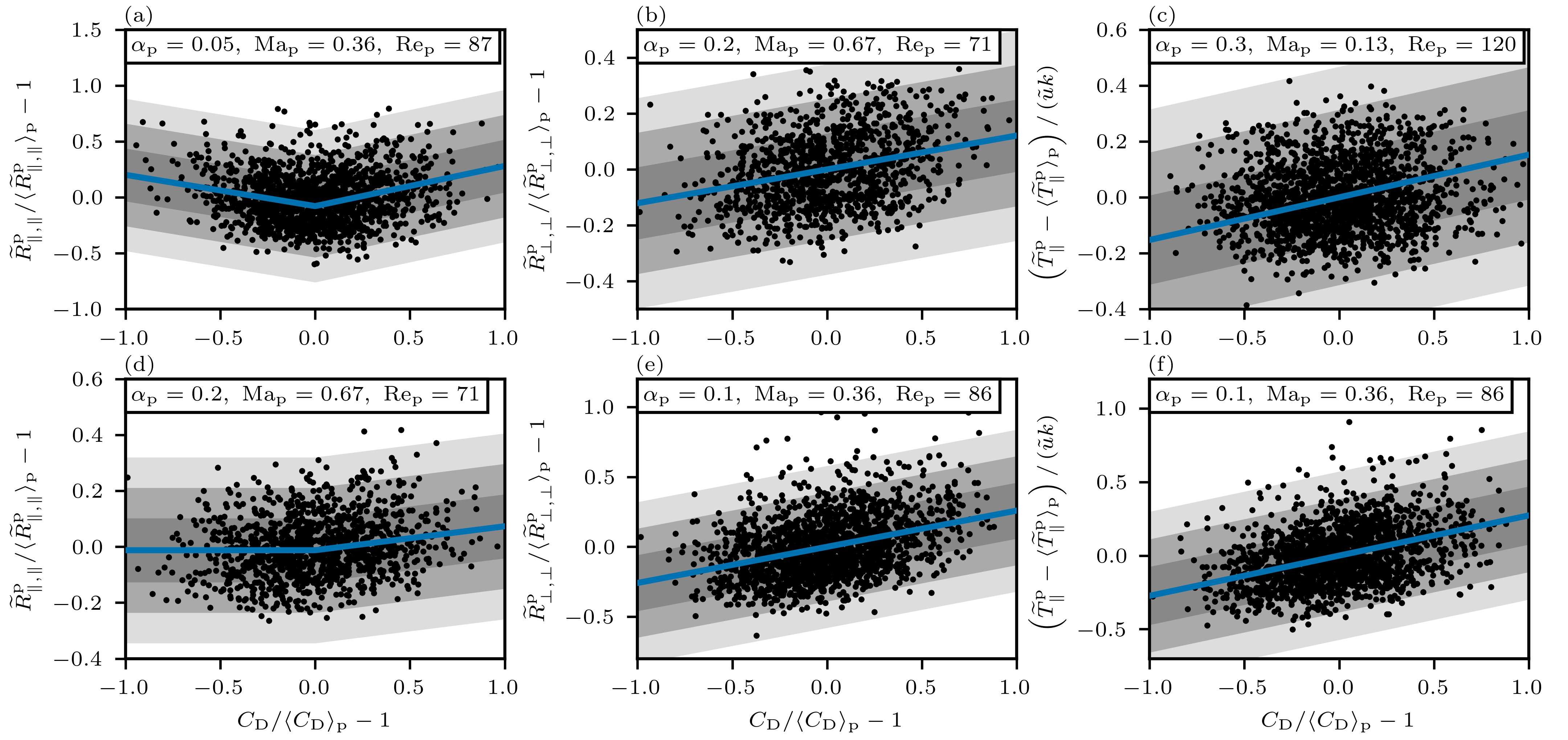}}
    \caption{Examples of particle-associated pseudo-turbulence plotted against $\Cd$. (a, d): $\Rparp$ , (b, e): $\Rperpp$, and (c, f): $\Tup$. The blue line and gray bands show the mean and one, two and three standard deviations predicted by \cref{eq:Rparpp} (a, b) and \cref{eq:Rperpp} (c, d), and by \cref{eq:Tup}: (e, f).} 
    \label{fig:Cd_R_corr}
\end{figure*}

\begin{table}
    \caption{Coefficients for the particle-associated pseudo-turbulence models, \cref{eq:Rparpp,eq:A1,eq:A2,eq:sparp,eq:Rperpp,eq:A3,eq:Tup,eq:A4,eq:sTp}.}
    \begin{tabular}{c|c|c|c|c|c}
    Parameter & Value & Parameter & Value & Parameter & Value\\
    \hline
    $F_1$ & -0.0022 & $G_1$ & -0.2867 & $H_1$ & 0.4992\\
    $F_2$ & -0.0219 & $G_2$ & 0.2176 & $H_2$ & -1.3528\\
    $F_3$ & 0.0932 & $G_3$ & 0.2826 & $H_3$ & -0.1358\\
    $F_4$ & -0.0135 & $G_4$ & -0.0644 & $H_4$ & -0.1463\\
    $F_5$ & 0.0361 & $G_5$ & 0.0466 & $H_5$ & 0.2583\\
    $F_6$ & 0.0403 & $G_6$ & 0.0973 & $H_6$ & -0.3339\\
    $F_7$ & -0.0761 & $G_7$ & -0.0081 & $H_7$ & -0.0407\\
    $F_8$ & 0.0599 & $G_8$ & -0.0235 & $H_8$ & -0.0806\\
    $F_9$ & 0.0164 &  &  &  & \\
    $F_{10}$ & 0.0453 &  &  &  & \\
    $F_{11}$ & -0.0265 &  &  &  & \\
    \end{tabular}
    \label{tab:FGH}
\end{table}
The $\Rpar - \Cd$ correlation scales differently with $\Cd'$ depending on the sign of $\Cd'$. We therefore model it as two linear branches where the slope in each region depends on $\alp$ and $\Map$. The model is
\begin{equation}
    \frac{\Rparp}{\pavg{\Rpar}} = 1 + A_1\left(\alp, \Map\right) + A_2\left(\alp, \Map\right)\frac{\Cd'}{\pavg{\Cd}} + \xi_\parallel(\alp, \Map),
    \label{eq:Rparpp}
\end{equation}
where $\pavg{\cdot}$ denotes the average over the particle ensemble, $\xi_\parallel$ is a normally distributed random variable with standard deviation $s_\parallel^\mathrm{p}\left(\alp, \Map\right)$, and
\begin{equation}
    A_1\left(\alp, \Map\right) = F_1/\left(\alp + F_2\right),
    \label{eq:A1}
\end{equation}
\begin{equation}
    A_2\left(\alp, \Map\right) = 
    \begin{cases}
    F_3 - 0.2 F_3/\left[\min(0.2,\ \alp)\right] & \text{if } \Cd' < 0\\
    F_4 + \frac{F_5}{\alp + F_6} + F_7\Map & \text{if } \Cd' \geq 0
    \end{cases},
    \label{eq:A2}
\end{equation}
\begin{equation}
    s_\parallel^\mathrm{p}\left(\alp, \Map\right) = F_8 + \frac{F_{9}}{\alp + F_{10}} + F_{11}\Map.
    \label{eq:sparp}
\end{equation}
The coefficients $F_1, F_2, \dots, F_{11}$ are reported in \cref{tab:FGH}. The model is plotted versus $\Cd$ along with the data from the simulations for two bulk flow conditions in \cref{fig:Cd_R_corr} (a, b).

The flow-normal component, $\Rperpp$ is modeled as
\begin{equation}
    \frac{\Rperpp}{\pavg{\Rperpp}} = 1 + A_3\left(\alp, \Map\right)\frac{\Cd'}{\pavg{\Cd}}+\xi_\perp^\mathrm{p}(\alp, \Rep),
    \label{eq:Rperpp}
\end{equation}
where $\xi_\perp^\mathrm{p}$ is a normally distributed variable with standard deviation $s_\perp^\mathrm{p}$, and 
\begin{equation}
     A_3\left(\alp, \Map\right) = G_1 + \frac{G_2}{\alp + G_3} + G_4\Map, \quad s_\perp^\mathrm{p}(\alp, \Rep) = \frac{G_5}{\alp + G_6} +\frac{ G_7\Rep}{300\alp} + G_8.
     \label{eq:A3}
\end{equation}
The coefficients $G_1, G_2, \dots, G_{8}$ are reported in \cref{tab:FGH}. The model is plotted versus $\Cd$ along with the data from the simulations for two bulk flow conditions  in \cref{fig:Cd_R_corr} (c, d).

Finally, $\Tup$ is modeled as
\begin{equation}
    \frac{\Tup - \pavg{\Tup}}{\fu k} = A_4\fulldep\frac{\Cd'}{\pavg{\Cd}} + \xi_{T}^\mathrm{p}\fulldep,
    \label{eq:Tup}
\end{equation}
where $\xi_{T}^\mathrm{p}$ is a normally distributed variable with standard deviation $s_T^\mathrm{p}$, and
\begin{equation}
    A_4\fulldep = H_1 + H_2\alp + H_3\Map + H_4\Rep/300,
    \label{eq:A4}
\end{equation}
\begin{equation}
    s_T^\mathrm{p}\fulldep=H_5 + H_6\alp + H_7\Map + H_8\Rep/300.
    \label{eq:sTp}
\end{equation}
The coefficients $H_1, H_2, \dots, H_{8}$ are reported in \cref{tab:FGH}, and the model is plotted versus $\Cd$ along with the data from the simulations for two bulk flow conditions in \cref{fig:Cd_R_corr} (e, f).

We note that in development of \cref{eq:Rparpp,eq:Rperpp,eq:Tup}, there was ambiguity in whether to include a dependence of $\Map$ or $\Rep$ for some of the terms. This could be avoided by using a larger dataset for the model development. However, the models are primarily dependent on $\alp$, with minor corrections due to $\Map$ and $\Rep$. It should also be noted that all the proposed Lagrangian models implicitly depend on all bulk flow parameters through the models for the mean values. 

\section{Concluding remarks}
In this study we have proposed an algebraic model for the pseudo-turbulent Reynolds stress tensor in compressible homogeneous flows through random, fixed, monodisperse particle clouds. The model was developed based on data from particle-resolved numerical simulations and is applicable for volume fractions, Reynolds numbers and Mach numbers in the ranges $\alp \in [0, 0.3]$, $\Map \in [0, 0.87]$ and $\Rep \in [30,266]$. It approaches the model of \citet{mehrabadi2015} in the incompressible limit. 

We have also developed a Lagrangian model for the local pseudo-turbulent Reynolds stress around a particle based on correlations of a local estimate of the Reynolds stresses and the drag force on the particle. This model can be employed in Euler-Lagrange simulations to obtain a more consistent application of force and Reynolds stress models.   

Additionally, corresponding models for the pseudo-turbulent transport of pseudo-turbulent kinetic energy are also introduced as a first effort aimed at closing the energy equation.

Python implementations of the models presented in this work are provided as supplementary material along with the pseudo-turbulence data used for calibration of the Eulerian models. 

\bibliographystyle{elsarticle-harv}
\bibliography{references}





\end{document}